\documentclass[letterpaper,twocolumn,10pt]{article}
\usepackage{usenix}
\usepackage{enumitem}

\usepackage{tikz}
\usepackage{amsmath}
\usepackage{amssymb}
\usepackage{filecontents}

\begin{document}
%-------------------------------------------------------------------------------

%don't want date printed
\date{}

\title{\Large \bf Kairos: Lightweight Testing Framework for Timing-Induced Interaction Failures \\
	in LTE and 5G Core Networks}

%for single author (just remove % characters)
\author{
	{\rm Wei Guo, Yuanhao Li, Hao Zheng, Junman Qin, Jun Kong, Jiapeng Li, Qiang Fu, Jiadai Wang, and Jiajia Liu}\\
	School of Cybersecurity, Northwestern Polytechnical University
}% end author
% copy the following lines to add more authors
% \and
% {\rm Name}\\
%Name Institution

\maketitle

%-------------------------------------------------------------------------------
\begin{abstract}
As cellular core networks evolve toward distributed and cloud-native architectures, control-plane interactions become more intricate and bring new challenges. 
%23
Among these challenges, we find that introducing specific timing between two control-plane interactions can cause network function crash, which we define as timing-induced interaction failures.
%30
Prior research primarily addresses identifying malformed inputs and specification violations, while timing-induced interaction failures remain largely unexplored. 
%20
In this paper, we conduct a systematic study of timing-induced interaction failures in LTE and 5G core networks. 
%20
First, we establish a taxonomy of control-plane interaction patterns and analyze the failure modes of each pattern. 
%20
Then, we design and implement Kairos, a lightweight testing framework to expose timing-induced interaction failures without analyzing cellular standard documents. 
%23
Evaluating Kairos on two open source and two commercial LTE and 5G core networks, we uncover 20 new vulnerabilities and reproduce 34 existing issues. 
%25
Our results show that timing-induced interaction failures are prevalent in LTE and 5G core networks and should be explicitly considered in future specifications.
%25

\end{abstract}

%-------------------------------------------------------------------------------
\section{Introduction}
\label{sec:introduction}
%-------------------------------------------------------------------------------
\nocite{3gpp:ts23401}
\nocite{3gpp:ts23501}
\nocite{3gpp:ts33501}
\nocite{3gpp:ts23502}
\nocite{3gpp:ts24501-r18}
\nocite{3gpp:ts33512}
\nocite{open5gs}
\nocite{free5gc}
\nocite{srsran}
\nocite{ueransim}
To accommodate the continuously growing traffic scale and diverse service demands, cellular core networks from LTE to 5G have gradually evolved toward distributed and cloud-native architectures\cite{3gpp:ts23401,3gpp:ts23501,3gpp:ts33501, sekigawa2022toward, mudvari2022exploring, dalgitsis2024cloud}.
In contrast to earlier monolithic designs, control-plane functions are now executed across multiple cooperating network functions through service-based interfaces.
In practice, control-plane network functions are commonly deployed as containerized services, decoupling their execution from fixed physical hosts.
This architectural evolution not only significantly improves LTE and 5G core network scalability, flexibility, and resource utilization efficiency, but also enables the system to better adapt to dynamic workloads and complex operational environments.

While the distributed and cloud-native architecture brings many benefits, it also introduces new risks in LTE and 5G core networks.
3GPP specifications define the request and response messages in control-plane interactions, and also establish the execution sequence of these interactions to implement specific procedures, such as UE registration and de-registration procedure\cite{3gpp:ts23502,3gpp:ts24501-r18,3gpp:ts33512}.
In practical systems, core networks process interactions according to the sequences defined by specifications, and allow multiple procedures to proceed concurrently.
However, the execution of interactions is subject to message transmission delays, queue-based processing, and network function execution latency.
When multiple procedures are processed concurrently, variability in execution timing can cause the realized execution order of interactions to deviate from the specification-defined execution sequence.
\textbf{We define such deviations caused by execution timing variability as timing-induced interaction anomalies}.
Notably, timing-induced interaction anomalies can arise during normal UE interactions without requiring any specifically crafted inputs.
These anomalies reflect abnormal interaction behaviors during execution,~\textbf{which may cause timing-induced interaction failures in core networks}.
For example, Sun et al. report that issuing an InitialUEMessage and a Registration Request at a specific timing can crash the AMF due to incorrect error handling in the GMM state machine \cite{sun20255gc}.
This naturally raises a key question: {\itshape when timing-induced interaction anomalies arise, what will happen in LTE and 5G core networks?}

\noindent\textbf{Prior Research.} Prior research propose a variety of techniques to identify security and implementation flaws in cellular core networks, and has been shown to be highly effective in uncovering failures caused by protocol semantics violations and malformed parameters\cite{sun20255gc, dong2025corecrisis, son2025citesting, al2024hermes, rahman2024cellularlint, wei2025unleashing, saifuzzaman2025dissecting, klischies2023instructions, hussain2021noncompliance, kim2019touching, chen2023sherlock, he2022intelligent, wang2023nlp}.
However, existing approaches exhibit two important limitations.
First, many prior approaches\cite{dong2025corecrisis, son2025citesting,al2024hermes, rahman2024cellularlint} analyze control-plane procedures using  state learning or focus on modeling the logical ordering of control-plane events as defined in specifications.
While effective for reasoning about protocol semantics, such approaches typically assume that interactions are realized at runtime in accordance with the intended logical order defined in the specifications.
Sun et al. can expose failures involving multiple control-plane procedures. 
Their approach primarily focuses on signaling sequence mutation\cite{sun20255gc}.
Overall, timing-induced interaction failures arising from deviations in execution sequences have not yet been systematically studied.
Second, many existing approaches are designed for a specific generation of cellular core networks and rely on analyzing   specific specifications\cite{dong2025corecrisis, hussain2021noncompliance, kim2019touching}.
This requires considerable analysis effort and hinders their adaptation to other core network generations(e.g., from LTE to 5G).

\noindent\textbf{Proposed approach.} In this paper, we conduct a systematic study of timing-induced interaction failures in LTE and 5G core networks. 
First, we characterize timing-induced interaction anomalies through three fundamental interaction patterns and analyze the potential failures associated with each pattern.
These patterns are defined based on the temporal relationships between concurrent control-plane procedures, including interleaving, nested, and incomplete interactions.
In addition, we design and implement Kairos (Kairos denotes the opportune moment, as opposed to arbitrary time, at which an event becomes consequential), a lightweight testing framework to expose failures caused by timing-induced interaction anomalies in LTE/5G core networks.
Kairos does not rely on analyzing protocol specifications.
It can be rapidly deployed across different generations of cellular core networks and efficiently discovers timing-induced interaction failures.

\noindent\textbf{Challenges.} To understand why timing-induced interaction anomalies arise, a key challenge lies in reasoning about how independent procedures interact when their executions are influenced by relative timing. Although each procedure is specified and implemented in isolation, multiple procedures may progress concurrently and interact through shared contexts at network functions. These interactions are neither explicitly constrained by protocol semantics nor explained by analyzing individual procedures in isolation. To address this challenge, we characterize control-plane interactions from an execution perspective, based on how interactions relate to each other in time, and separate the effects into three interaction patterns: interleaving, nested, and incomplete.

To systematically explore timing-induced interaction failures in LTE and 5G core networks, we identify three key challenges that must be considered before designing Kairos.
First, timing-induced interaction anomalies often manifest only within narrow and transient execution windows. They are difficult to trigger and observe through conventional testing approaches.
Second, interactions among concurrently executing control-plane procedures are inherently complex.
It is difficult to construct a comprehensive model that captures all possible control-plane interactions.
Third, cellular core networks exhibit substantial architectural and implementation heterogeneity across generations.
These differences make it difficult to apply a unified analysis of timing-induced interaction failures in different core networks.
To address these challenges, Kairos is designed with three modules: the UE Behavior Pattern Generator, the Timing Scenario Orchestrator, and the Network Impairment Engine.
These modules enable Kairos to systematically perturb execution timing and increase the likelihood of observing timing-induced interaction failures in both LTE and 5G core networks.

\noindent\textbf{Findings.} We evaluate Kairos on two open-source and two commercial cellular core networks to assess its effectiveness in exposing timing-induced interaction failures.
We discover 20 new vulnerabilities and successfully reproduce 34 existing issues. 
These vulnerabilities arise across different UE behaviors, involved network functions, and vendors, spanning both LTE and 5G deployments.
Among these vulnerabilities, 11 have been assigned CVE identifiers, while the remaining issues have been responsibly disclosed to the open-source maintainers or commercial vendors.

{\noindent\textbf{Contribution.} We make the following contributions:
	\begin{itemize}[leftmargin=*, topsep=0pt, itemsep=0pt, parsep=0pt, partopsep=0pt]
		\item We present a systematic study of timing-induced interaction failures in LTE and 5G core networks.
		To the best of our knowledge, this is the first work to characterize such failures through a unified taxonomy.
		
		\item We design and implement Kairos, a lightweight testing framework for exposing timing-induced interaction failures.
		Kairos does not rely on analyzing cellular standards and can be readily applied to both LTE and 5G core networks.
		
		\item Applying Kairos to two open-source and two commercial cellular core networks, we identify 20 new vulnerabilities and reproduce 34 existing issues.
		
		\item We release Kairos as an open-source testing framework to facilitate future research on studying timing-induced interaction failures. The source code is available at \url{https://anonymous.4open.science/r/Kairos-FC84/Readme.md}
		
	\end{itemize}
}
%-------------------------------------------------------------------------------

\section{Background}
\label{sec:background}
%-------------------------------------------------------------------------------
\noindent\textbf{Control-Plane Procedures and Interactions.} Cellular core networks provide essential control-plane services through a set of control-plane procedures, including user registration, mobility management, and session management\cite{3gpp:ts23502, 3gpp:ts33501}.
In 3GPP specifications, a control-plane procedure is defined as a process composed of multiple well-formed interactions.
A procedure consists of multiple processing steps and is advanced through interactions among network functions, with its execution potentially spanning multiple network functions.
An interaction is typically realized as a pair of request--response messages (e.g., HTTP/2 in 5G SBA), whose outcomes determine subsequent execution path of the procedure. 

\noindent\textbf{Control-Plane Execution and Scheduling.} As cellular core networks evolve from LTE to 5G, control-plane functionalities become increasingly decoupled and the types of network functions continue to grow.
In this setting, interactions that advance control-plane procedures are typically executed asynchronously, and control-plane messages are subject to queuing and scheduling at individual network functions.
In practice, cellular core networks are gradually adopting cloud-native architectures.
Control-plane network functions are commonly instantiated as containerized services and may be deployed across different physical hosts or data centers.
Under such deployment environments, the execution of control-plane procedures relies on the transmission of control-plane messages over the network, as well as their reception and processing at the corresponding network functions.
As a result, the intended ordering of procedure execution defined by specifications may not be preserved at runtime.

\begin{figure}[t]
	\centering
	\includegraphics[width=\columnwidth]{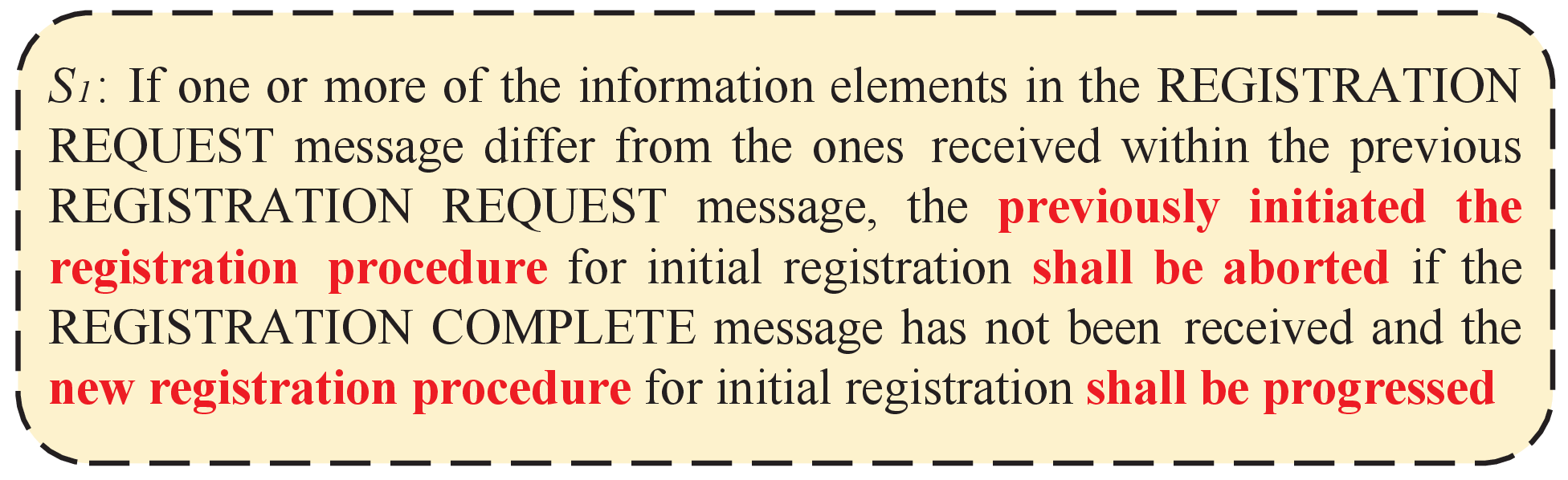}
	\caption{An abnormal registration handling case specified in Section 5.5.1.2.8 of TS 24.501.}
	\label{fig:protocol}
\end{figure}

\noindent\textbf{Protocol Semantics of Conflicting Procedures.} In addition to specifying the expected behavior of control-plane procedures under normal conditions, the 3GPP specifications also define handling semantics for abnormal cases\cite{3gpp:ts24501-r18}.
As shown in Figure~\ref{fig:protocol}, Section~5.5.1.2.8 of 3GPP TS~24.501 specifies that if a new registration request is received before the current registration procedure completes, the network shall abort the current procedure and initiate a new one.
Such rules define the intended priority and handling order among concurrent or repeated procedure instances.
Similar semantics are specified for other control-plane procedures.
However, these specifications mainly define that the ongoing procedure should be aborted in favor of the new one , when multiple procedure instances conflict.
In practice, whether and when this rule can be enforced depends on the execution of asynchronous request--response interactions across network functions.
This execution gap may introduce subtle risks and lead to unexpected network function failures.
%-------------------------------------------------------------------------------
\section{Taxonomy of Interaction Patterns}
\label{sec:taxonomy}
%-------------------------------------------------------------------------------
\subsection{Scope and Modeling Conditions}
We study timing-induced interaction anomalies defined in Section~\ref{sec:introduction} and focus on how control-plane interactions from different procedures interfere with each other under specific timing conditions.\textbf{We adopt the following modeling conditions to establish a minimal and tractable execution setting}.
First, each control-plane procedure executes according to the interaction sequence defined by the specifications and is associated with the same UE.
Second, each control-plane interaction consists of a well-formed request--response message pair.
Third, request and response messages are not delivered instantaneously, but are subject to transmission and processing delays.
Fourth, network functions can initiate a new control-plane request without synchronously waiting for the response to a previously issued request.
Under these conditions, we characterize timing-induced interaction anomalies by examining the temporal relationships among concurrent control-plane interactions.
Interactions that do not satisfy these conditions may still involve concurrent processing at other network functions; however, such cases are outside the scope of this work.

Without loss of generality, we consider a minimal scenario involving two concurrent interactions associated with the same UE, as illustrated by the normal case in Figure~\ref{fig:Interactions}(a). Two concurrent interactions are treated as the basic analysis unit, since additional interactions are decomposed into pairwise execution relationships, without introducing new temporal structures. Such two interactions span at most three network functions, because it is enough to capture shared contexts and cross-procedure dependencies. Within this execution model, we identify three fundamental control-plane interaction patterns: interleaving, nested, and incomplete interactions.

\subsection{Interleaving Interactions}
An interleaving interaction pattern arises when two independent control-plane interactions partially overlap in execution at the same network function.
Before receiving the response to the first interaction, the network function may initiate and advance the second interaction.
Consequently, the response of the first interaction may be processed after the second request.
The defining characteristic of interleaving interactions is that both interactions make progress concurrently at the same network function, and their request--response pairs are processed in an interwoven order.

\begin{figure*}[t]
	\centering
	\includegraphics[width=\textwidth]{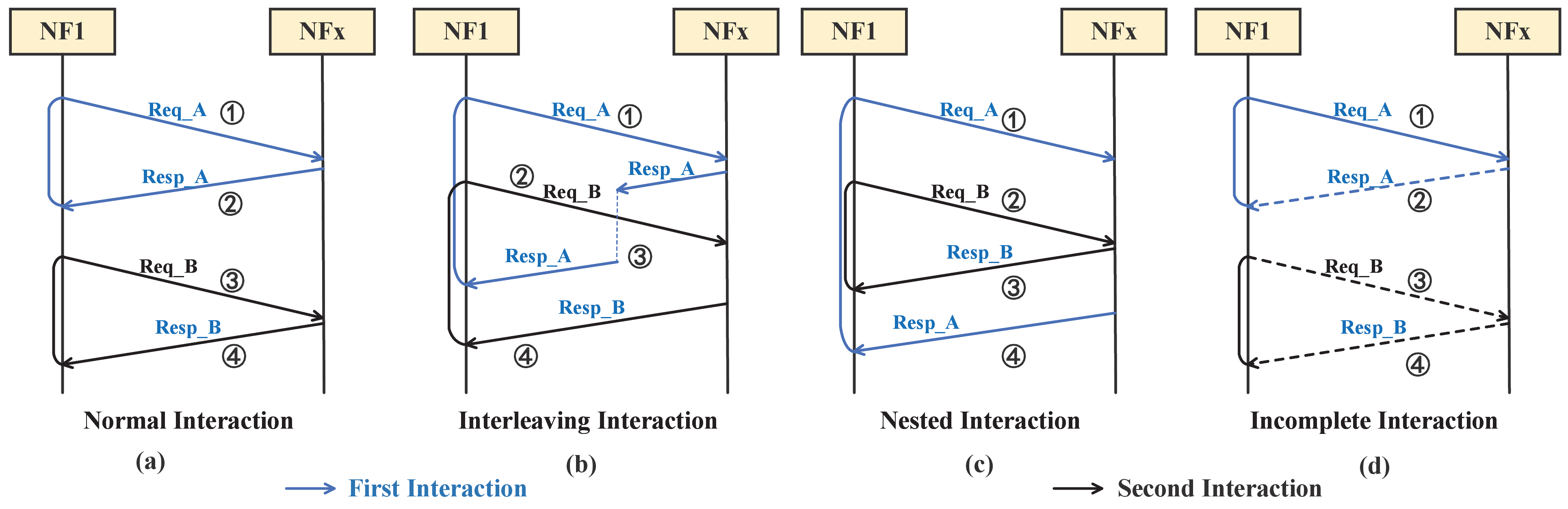}
	\caption{Illustration of interaction patterns involving two concurrent control-plane interactions.
	Each interaction consists of a request--response message pair.
	(a) A normal interaction pattern, where two interactions follow the execution order defined by specifications.
	(b)--(d) Timing-induced interaction anomalies, including interleaving, nested, and incomplete interaction patterns}
	\label{fig:Interactions}
\end{figure*}

Figure~\ref{fig:Interactions}~(b) illustrates an interleaving interaction pattern. \textcircled{1}The network function issues~\texttt{Req\_A}~for the first interaction. \textcircled{2}Then the network function initiates the second interaction by issuing~\texttt{Req\_B}. 
\textcircled{3}\texttt{Resp\_A}~is processed after the second interaction has been initiated, leading to an interleaved request--response ordering at runtime. 
\textcircled{4}After that,~\texttt{Resp\_B}~is processed.
Such interleaving behavior is relatively easy to trigger, when two procedures involve interactions initiated by the same network function.

This pattern is risky because it can break \emph{state consistency} across overlapping procedures within a network function.
In cellular core networks, control-plane procedures are typically implemented as event-driven handlers.
As control-plane messages arrive, these handlers update shared per-UE control state, including procedure phases and security context status.
When two interleaving interactions \textbf{do not operate on shared per-UE state}, their execution behavior remains equivalent to the normal case: responses are processed in the order expected by their corresponding requests.
However, when multiple interactions \textbf{operate on shared per-UE state}, interleaving can become problematic.
When the delayed response from the first interaction arrives, it may be handled under a state that has already been updated by the second interaction.
As a result, interleaving can cause network functions to process responses under unexpected state transitions that may eventually manifest as network function failure.

%-------------------------------------------------------------------------------

\subsection{Nested Interactions}
A nested interaction pattern arises when one control-plane interaction is initiated and executed entirely within the execution interval of another interaction at the same network function.
Before receiving the response to the first interaction, the network function may initiate the second control-plane interaction and allow it to completely execute.
As a result, the second interaction completes its execution before the first interaction finishes.
The defining characteristic of nested interactions is that one interaction fully completes before the outer interaction resumes execution, forming a strict containment relationship at runtime.

Figure~\ref{fig:Interactions}~(c) illustrates a nested interaction pattern.
\textcircled{1}The network function issues~\texttt{Req\_A}~for the first interaction. \textcircled{2}Before receiving~\texttt{Resp\_A}, the network function initiates the second interaction by issuing~\texttt{Req\_B}.
\textcircled{3}The second interaction receives ~\texttt{Resp\_B} and completes its execution. \textcircled{4}After that, the network function receives~\texttt{Resp\_A}.
As a result, the execution of the second interaction is fully contained within the execution interval of the first interaction at runtime.

This pattern is risky because it can lead to \emph{inconsistent resource usage} across concurrent control-plane procedures within a network function.
In cellular core networks, different control-plane procedures may independently access and manipulate shared resources associated with the same UE, such as UE identifiers, context handles, and session-related objects.
When two nested interactions \textbf{belong to the same control-plane procedure}, their execution typically follows a well-defined resource lifecycle.
Resource updates and releases are coordinated within the procedure.
However, when two nested interactions \textbf{do not belong to the same control-plane procedure}, the second interaction modifies or releases the shared resources before the first interaction finishes, the first interaction may subsequently fail to access or operate on the expected resources.
As a result, nested interactions can cause network functions to fail when accessing required resources, and may eventually manifest as network function failures.

\subsection{Incomplete Interactions}
An incomplete interaction pattern arises when a control-plane interaction is initiated but does not reach a request--response closure at a network function.
After issuing a request from the first interaction, the network function does not observe or process the corresponding response, and the first interaction terminates without completing its execution lifecycle.
Under such conditions, the execution of subsequent control-plane interactions becomes uncertain.
A second interaction may still be initiated and complete normally. 
It may fail during initiation or response processing due to the incomplete execution of the first interaction.
As a result, the interaction may fail to establish the execution context expected by subsequent control-plane processing at runtime.

Figure~\ref{fig:Interactions}~(d) illustrates an incomplete interaction pattern.
\textcircled{1}The network function issues~\texttt{Req\_A}~for the first interaction.
\textcircled{2}The corresponding response~\texttt{Resp\_A}~is not successfully received or processed, leaving the first interaction incomplete.
\textcircled{3}Under such conditions, the network function may initiate the second interaction by issuing~\texttt{Req\_B}.
\textcircled{4}Even if~\texttt{Req\_B}~is successfully issued, the network function may fail to process~\texttt{Resp\_B}.
As a result, the first interaction does not reach a request--response closure, and the expected execution context is not fully established at runtime.

This pattern is risky because it introduces a \emph{mismatch between logical decisions and execution outcomes} within a network function.
In cellular core networks, control-plane executions may assume that an initiated interaction will eventually complete.
For example, when a network function issues several requests to another network function within a control-plane procedure, validation may be performed only for the initial interaction, with subsequent interactions proceeding under the conditions of eventual completion.
As a result, incomplete interactions can cause control-plane processing to proceed based on incomplete or missing execution contexts, and may eventually manifest as network function failures.

\section{Problem Setting and Approach Overview}
\label{sec:problem-setting}
In this section, we focus on how to systematically expose timing-induced interaction failures in LTE and 5G core networks.
First, we define the threat model in this work. 
Then, we discuss the challenges that make timing-induced interaction failures difficult to expose in practice.
Finally, we provide an overview of our approach.

\subsection{Threat Model}
\noindent\textbf{Adversary.} 
The adversary is a legitimate UE attached or registered to LTE and 5G core networks.
The UE is assigned a single valid subscription identity and complies with the specification.
Specifically, the adversary does not spoof other subscribers, does not introduce abnormal parameters, and does not compromise or bypass any cryptographic mechanisms.
The adversary also does not control or modify any network components, including base stations (eNB/gNB) or network functions.

\noindent\textbf{Capabilities.}
The adversary capabilities are limited to initiating control-plane procedures that are permitted for a normal UE.
These procedures may include, but are not limited to, repeated registration requests, combinations of registration and de-registration procedures, and session establishment procedures.
Multiple control-plane procedures initiated by the UE are semantically independent and valid in isolation, and their execution reflects realistic scenarios such as user mobility, intermittent connectivity, or bursty access behavior.
The adversary does not control internal network scheduling, message processing order, or per-message transmission delays.

\noindent\textbf{Execution Environment.}
We assume that control-plane messages are transmitted and processed asynchronously and concurrently in the target cellular core network.
Such behaviors naturally arise in operational networks due to asynchronous message delivery, queue-based processing, resource contention, and load fluctuations.
In our experiments, we emulate such timing variations by introducing external timing perturbations across different network components, such as the AMF and SMF.

\noindent\textbf{System Inputs and Goal.}
As shown in capabilities, the system inputs consist of UE-initiated control-plane procedures and their combinations.
Instead of injecting malformed messages, Kairos perturbs the execution timing of interactions to induce variations in execution ordering and completion among multiple procedures associated with the same UE.
Our goal under this threat model is to determine whether such timing-induced anomalies can expose unintended network function failures.

\subsection{Challenges and Design Rationales}
\label{subsec:Challenges and Design Rationales}
\noindent\textbf{Challenge 1: Narrow and Transient Execution Windows.}
Timing-induced interaction anomalies often manifest only within extremely narrow and transient execution windows.
In cellular core networks, individual interactions are typically short-lived, with request--response executions completing within a few milliseconds.
Consequently, timing-induced interaction anomalies can arise only within an execution window that is much narrower than the execution window of an individual interaction.
Nevertheless, given the large subscriber population and the continuous running of core networks, these low-probability timing-induced interaction anomalies can still accumulate into a practical and persistent risk, leading to network function failures.

\noindent\textbf{Design Rationale 1.} 
To address the narrow and transient execution windows described above,
Kairos amplifies the execution conditions under which timing-induced interaction anomalies can arise through two mechanisms.
First, Kairos introduces controlled delays to control-plane message delivery, extending the execution window of individual interactions.
This effectively enlarges the execution window in which timing-induced interaction anomalies can arise.
Second, Kairos increases the frequency and concurrency of UE-initiated control-plane procedures.
This raises execution pressure within network functions and increases the likelihood that multiple procedures progress under closely spaced execution timing.

\noindent\textbf{Challenge 2: Complex Timing-Induced Interactions.}
Timing-induced interaction anomalies can be explained through the relative execution timing of individual control-plane interactions.
However, discovering such anomalies in practice is considerably more challenging.
In operational core networks, a large number of control-plane procedures may execute concurrently across multiple network functions.
Although the interactions themselves are well defined by specifications, their executions can combine in complex ways at runtime.
As the number of concurrent procedures, involved network functions, and execution dependencies increases, timing-induced interactions are shaped by complex combinations of execution timing, making them difficult to anticipate or reason about systematically.

\begin{figure*}[t]
	\centering
	\includegraphics[width=0.95\textwidth]{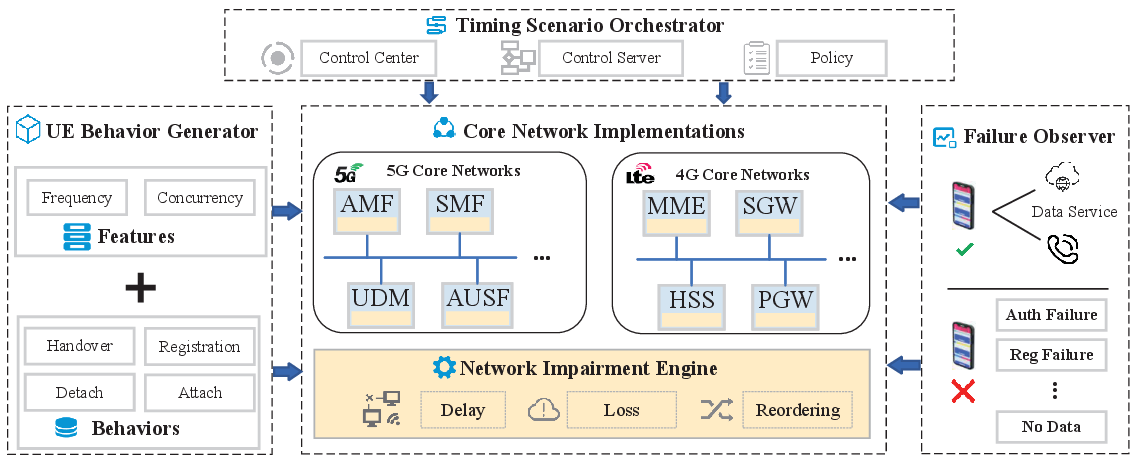}
	\caption{Design of Kairos.}
	\label{fig:system}
\end{figure*}

\noindent\textbf{Design Rationale 2.}
Although such interactions arise from complex combinations of execution timing across multiple procedures, they ultimately manifest through message exchanges between network functions.
Kairos therefore avoids orchestrating timing at individual procedures or interactions.
Instead, it selectively perturbs execution ordering along particular control-plane execution paths rather than uniformly perturbing the entire system.
By configuring timing perturbations across network functions in a unified manner, Kairos does not require any modification to network function implementations.
Such targeted perturbations can simultaneously affect multiple procedures and interactions that traverse the perturbed network functions, enabling Kairos to intentionally trigger timing-induced interaction anomalies.
This design makes Kairos lightweight and scalable, while also reducing noise from indiscriminate timing changes and facilitating the attribution of observed anomalous behaviors to concrete control-plane interactions.

\noindent\textbf{Challenge 3: Heterogeneity Across Core Network Generations and Implementations.}
Cellular core networks exhibit significant heterogeneity across generations, architectures, and implementations.
For example, LTE and 5G core networks differ substantially in their control-plane decomposition, interface abstractions, and interaction granularity, even when supporting procedures of similar functions.
As a result, analysis and testing approaches that rely on protocol-specific assumptions or detailed modeling of control-plane logic are often tightly coupled to a particular generation or implementation.
When protocols evolve or system architectures change, such approaches typically require substantial re-engineering, which limits their applicability and increases the cost of migration across heterogeneous core networks.
This heterogeneity poses a fundamental challenge to systematically exposing timing-induced interaction failures in both LTE and 5G core networks using a unified and lightweight approach.

\noindent\textbf{Design Rationale 3.}
To address heterogeneity across core network generations and implementations, Kairos adopts a lightweight and general testing approach.
Instead of relying on specification analysis, it drives testing through UE-initiated control-plane procedures and perturbs execution timing externally.
Because all cellular core networks, including both LTE and 5G systems, must process UE-initiated procedures and exchange control-plane messages among network functions. 
This design naturally generalizes across different generations and implementations.
Kairos therefore requires minimal adaptation when migrating across heterogeneous core networks.
The same UE behaviors and timing perturbation mechanisms can be reused across different systems, while differences manifest in execution behaviors and failure modes.

\begin{table*}[t]
	\centering
	\caption{UE Behaviors Primitives Across Cellular Generations}
	\label{tab:ue-access-patterns}
	\small
	\begin{tabular}{ccccc}
		\hline
		\textbf{UE Behaviors} &
		\textbf{LTE} &
		\textbf{5G} &
		\textbf{Scope} &
		\textbf{Typical Real-World Scenarios} \\
		\hline
		Registration &
		-- & \checkmark &
		UE context &
		UE power-on, turning off airplane mode, entering network coverage \\
		
		Attach &
		\checkmark & -- &
		UE context &
		Initial LTE network access after power-on or SIM re-insertion  \\
		
		Handover &
		-- & \checkmark &
		UE context &
		Cell change during driving or on high-speed trains \\

		De-registration &
		-- & \checkmark &
		UE context &
		UE power-off, user-initiated disconnect, or temporary loss of coverage \\
		
		Detach &
		\checkmark & -- &
		UE context &
		UE power-off, network-triggered release, or core-side context cleanup \\
		
		Service request &
		-- & \checkmark &
		Session context &
		Restore data connectivity after idle period or paging response \\
		
		Session establishment &
		-- & \checkmark &
		Session context &
		Initiating data services for applications such as web or video access  \\
		
		Session release &
		-- & \checkmark &
		Session context &
		Release active data session due to user action or network policy \\
		\hline
	\end{tabular}
\end{table*}

\subsection{Approach Overview}
We design and implement Kairos, a lightweight testing framework, which can expose timing-induced interaction failures by perturbing execution timing under specification-defined UE behavior.
As shown in Figure~\ref{fig:system}, Kairos consists of four main components: the UE Behavior Generator, the Timing Scenario Orchestrator, the Network Impairment Engine, and the Failure Observer.
Unlike existing approaches that rely on specification analysis or malformed message injection, Kairos operates entirely within legitimate UE-initiated control-plane procedures.
Rather than attempting to precisely control message ordering or internal scheduling decisions, Kairos selectively perturbs execution ordering along specific control-plane execution paths.
By enlarging otherwise narrow execution windows, Kairos increases the likelihood that independent control-plane procedures overlap during execution.
This approach enables Kairos to expose failures that arise from timing-induced interaction anomalies.

\section{Design of Kairos}
In this section, we present the detailed design of Kairos, focusing on how its components work together to identify timing-induced interaction failures.

\subsection{System Architecture}
Kairos is designed as a lightweight testing framework that exposes timing-induced interaction failures by selectively perturbing execution ordering under specification-defined UE behavior.
The UE Behavior Generator defines which control-plane behaviors are initiated by UEs.
It captures how these behaviors are invoked over time, including their invocation frequency and degree of concurrency.
Given these inputs, the Timing Scenario Orchestrator determines which parts of the control-plane execution are subject to timing perturbation.
It then defines timing perturbation policies that govern how execution ordering is altered at selected network boundaries.
The Network Impairment Engine enforces these policies by impairing request--response message exchanges between network functions.
It introduces controlled delay, packet loss, and delivery-time variation to perturb execution timing.
Finally, the Failure Observer detects failures by monitoring whether a UE can successfully complete access procedures.
This modular structure cleanly separates input generation, execution-time perturbation, and behavior observation.
As a result, Kairos can be readily applied across different generations and implementations of cellular core networks without depending on protocol versions or architectural details.

%-------------------------------------------------------------------------------

\subsection{UE Behavior Generator}
The UE Behavior Generator defines the input space from the perspective of a legitimate user equipment.
Rather than operating on individual protocol messages or internal state transitions, the UE Behavior Generator treats UE-initiated control-plane procedures as its basic input units.
Each procedure is specification-defined and valid in isolation.
It can be initiated by a normal UE without requiring any knowledge of the core network’s internal execution state.
By adopting this UE-centric abstraction, the UE Behavior Generator decouples input generation from protocol-specific message formats and implementation details.

Based on this UE-centric abstraction, the UE Behavior Generator considers a representative set of UE-initiated control-plane procedures that commonly occur in operational cellular networks.
These procedures span both LTE and 5G core networks and capture essential UE behaviors related to access, mobility, and session management. Table~\ref{tab:ue-access-patterns} summarizes the procedure primitives considered in this work, along with their applicability across cellular generations and typical real-world scenarios.
The UE Behavior Generator focuses on a core set of procedure primitives that are sufficient to exercise diverse execution paths in the control plane.

Beyond individual procedure primitives, the UE Behavior Generator focuses on how legitimate UE behaviors are invoked over time.
The UE Behavior Generator models their temporal invocation patterns, such as invocation frequency, concurrency, and composition.
A single procedure may be triggered repeatedly within short intervals, multiple independent procedures may be initiated concurrently, and different procedures may overlap during execution.
Each such behavior remains specification-defined and valid in isolation. Timing effects arise solely from how these behaviors are scheduled and combined over time.

\subsection{Timing Scenario Orchestrator}
The Timing Scenario Orchestrator constructs execution-time scenarios that shape control-plane execution conditions in the cellular core network.
It does not directly control the processing of individual protocol procedures or messages.
Instead, it determines where timing perturbations are applied and over which execution intervals they take effect.
By selecting specific network functions or communication paths and associating them with a timing scenario, the orchestrator influences how control-plane interactions progress during execution.
These perturbations do not modify protocol logic, message contents, or control-plane semantics.
Rather, they indirectly affect the relative progress of concurrent procedures by altering execution conditions at selected points in the network.

\begin{figure}[t]
	\centering
	\includegraphics[width=0.9\columnwidth]{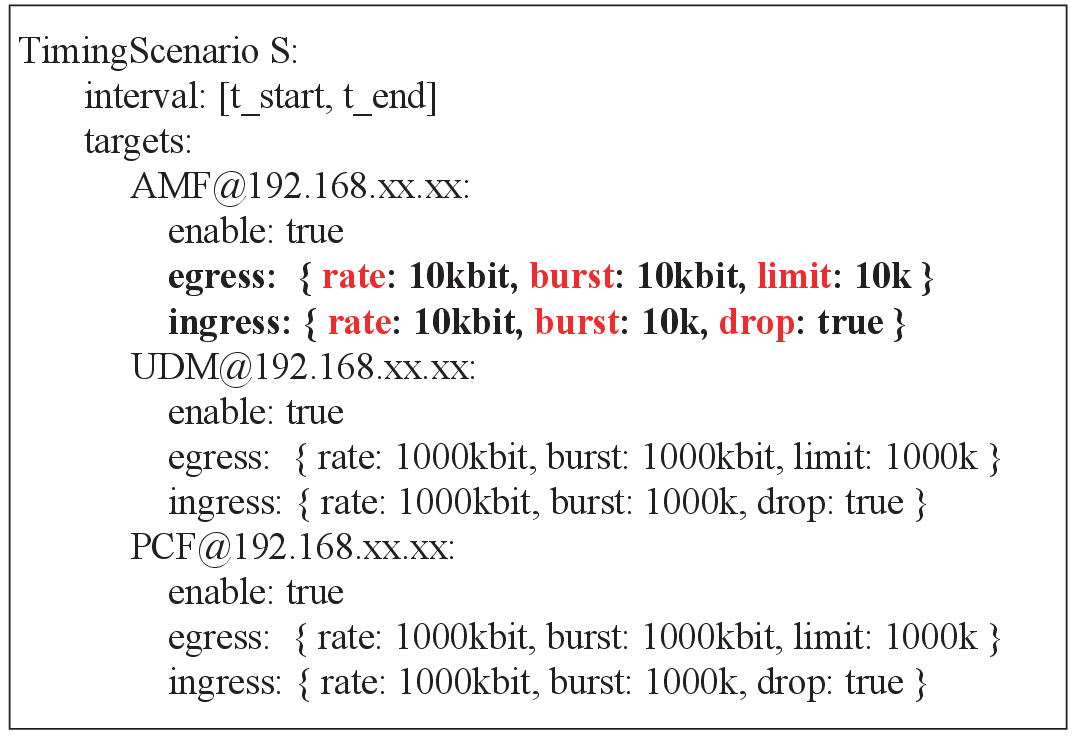}
	\caption{An example timing scenario configuration. The orchestrator selects network functions and specifies ingress/egress perturbation parameters over an execution interval.}
	\label{fig:or}
\end{figure}

Figure~\ref{fig:or} illustrates the execution view adopted by the Timing Scenario Orchestrator.
In each timing scenario, the orchestrator selects a subset of network functions and specifies timing perturbations on their communication interfaces.
These perturbations are configured independently for ingress and egress directions and are defined using network-level parameters, such as rate limits, burst sizes, and packet dropping.
By varying which network functions are affected and how their interfaces are configured, the orchestrator creates different execution conditions.
For example, a timing scenario may enable bandwidth limiting and packet dropping on the ingress and egress interfaces of the AMF and UDM, while applying milder perturbations to other network functions such as the SMF or AUSF.
Under such a scenario, identical UE behaviors may lead to different degrees of overlap among concurrent control-plane procedures.
This allows timing interactions among procedures to be observed, while all interactions remain specification-defined and without relying on internal control-plane implementations.

The Timing Scenario Orchestrator is deployment-agnostic and supports both single-node and multi-node cellular core network deployments.
It adopts a centralized orchestration model, in which a central controller disseminates timing scenario configurations to target core-network hosts.
Each host continuously listens for orchestration commands and locally invokes the Network Impairment Engine to enforce the specified execution conditions.
As a result, the same orchestration can be reused across heterogeneous deployments, while consistently shaping the temporal relationships among concurrently executing control-plane procedures.

\subsection{Network Impairment Engine}
The Network Impairment Engine realizes timing scenarios defined by the Timing Scenario Orchestrator by translating timing variations and concurrent execution into concrete and reproducible execution conditions at runtime.
Rather than modifying the core network or individual components, Kairos perturbs execution timing by introducing controlled impairments along communication paths.
This allows interactions defined by cellular specifications to proceed under altered timing conditions.

Cellular core networks are complex and heterogeneous.
Timing-induced interaction failures often arise from how interactions are delivered and processed across network functions, rather than from the internal behavior of a single component.
By operating on communication paths instead of internal execution, the Network Impairment Engine affects interaction timing without relying on protocol semantics.
The Network Impairment Engine supports multiple impairment conditions that jointly shape execution timing, including controllable transmission delay, packet loss, and sustained congestion introduced through bandwidth limitation and queue buildup.

\subsection{Failure Observer}
Kairos relies on the Failure Observer to determine whether a timing scenario exposes a failure in the cellular core network.
The Failure Observer is defined based on externally observable signals and does not inspect internal logs or protocol states.
We consider two classes of failures.
First, core network crashes, where a network function process terminates unexpectedly or becomes unreachable at its service interface.
Second, persistent service unavailability, where specification-defined UE requests repeatedly fail over a sustained period, preventing essential procedures such as registration or session establishment from completing successfully.
These failures are externally visible and directly impact service continuity.

The Failure Observer is conservative by design and aligned with our study objectives.
It does not attempt to detect silent bugs, transient inconsistencies, or internal state corruption that does not immediately affect service behavior.
Instead, it focuses on failures that are high-impact, reproducible, and clearly attributable to timing-induced execution interactions.
By relying on external symptoms rather than internal assumptions, the Failure Observer avoids overfitting to specific implementations and enables consistent failure judgment across heterogeneous core networks.
This conservative design prioritizes reliability and interpretability while still capturing the most consequential timing-induced interaction failures.
%-------------------------------------------------------------------------------
\section{Implementation}
%-------------------------------------------------------------------------------
We implement Kairos on top of open-source cellular core network implementations, including open5gs\cite{open5gs} and free5gc\cite{free5gc}, and further validate it on two commercial cellular core networks.
All experiments are conducted in containerized environments.
As a result, the same experimental setup can be applied across both single-node and distributed core network deployments.
To control execution timing, we employ a small set of representative timing configurations rather than fine-grained continuous delay sweeps.
We focus on discrete timing points that lead to qualitatively different timing-induced interaction failures.
We adopt a conservative failure definition based solely on externally observable UE outcomes, such as unsuccessful registration, failed session establishment, or sustained service unavailability.
Our evaluation does not rely on internal logs or implementation-specific states, nor does it attempt to classify failure causes beyond their externally observable impact.

%-------------------------------------------------------------------------------
\section{Evaluation}
\label{sec:Evaluation}
%-------------------------------------------------------------------------------
With the following research questions, we evaluate Kairos for identifying timing-induced interaction failures in cellular core networks.
\begin{itemize}[leftmargin=*, topsep=0pt, itemsep=0pt, parsep=0pt, partopsep=0pt]
	
	\item \textbf{RQ1.}
	How general is Kairos for identifying timing-induced interaction failures in LTE and 5G core networks? (§7.1)
	
	\item \textbf{RQ2.}
	How effectively can Kairos identify timing-induced interaction failures across different interaction patterns? (§7.2)
	
	\item \textbf{RQ3.}
	What is the efficiency and deployment overhead of Kairos as a lightweight testing framework in identifying timing-induced interaction failures? (§7.3)
	
\end{itemize}

\begin{table*}[t]
	\centering
	\caption{New timing-induced vulnerabilities identified by Kairos.}
	\label{tab:vulnerability-overview}
	\setlength{\tabcolsep}{5pt}
	\scriptsize
	\begin{tabular}{|l|l|l|l|l|l|l|}
		\hline
		\textbf{Flaw} &
		\textbf{Target} &
		\textbf{Comp.} &
		\textbf{UE Behaviors} &
		\textbf{Interaction Pattern} &
		\textbf{CVE ID} &
		\textbf{Root cause} \\
		\hline
		C1 & open5gs & AMF &
		HO $||$ Reg &
		Interleaving &
		CVE-2025-1893 &
		Received Nudm-SDM response in authentication state \\
		\hline
		C2 & open5gs & AMF &
		Reg $||$ Reg &
		Interleaving &
		CVE-2025-1925 &
		Received a duplicated SM-context update in authentication state \\
		\hline
		C3 & open5gs & AMF &
		Reg $||$ Reg &
		Interleaving &
		CVE-2025-25774 &
		Received a mismatched Policy-control response in authentication state \\
		\hline
		C4 & open5gs & AMF &
		Reg $||$ Reg &
		Interleaving &
		CVE-2025-52288 &
		Received a Downlink NAS transport in registered state \\
		\hline
		C5 & open5gs & AMF &
		Reg $||$ Reg &
		Interleaving &
		CVE-2025-8799 &
		Received a Nudm-SDM response in initial-context-setup state \\
		\hline
		C6 & open5gs & AMF &
		Reg $||$ Reg &
		Interleaving &
		CVE-2025-8801 &
		Received a Nudm-SDM response in exception state \\
		\hline
		C7 & open5gs & AMF &
		Reg $||$ Reg &
		Interleaving &
		CVE-2025-8803 &
		Received a Policy-control response in security-mode state \\
		\hline
		C8 & open5gs & AMF &
		Reg $||$ Reg &
		Interleaving &
		CVE-2025-8804 &
		Received an Nsmf-pdusession response in authentication state \\
		\hline
		C9 & open5gs & AMF &
		Reg $||$ Reg &
		Interleaving &
		CVE-2025-9405 &
		Received a Nudm-UECM response in exception state \\	
		\hline
		C10 & open5gs & SMF &
		Reg $||$ Reg &
		Interleaving &
		CVE-2025-8802 &
		Received a Nudm-UECM response in SMF operational state \\
		\hline
		C11 & open5gs & SMF &
		Reg $||$ Reg &
		Interleaving &
		CVE-2025-8805 &
		Received a Namf-comm response in wait-pfcp-deletion state \\
		\hline
		C12 & open5gs & MME &
		Att $||$ Att &
		Interleaving &
		requested &
		Received an eNB-UE context during ESM handling in inactive state \\
		\hline
		C13 & open5gs & SGW-C &
		Att $||$ Att &
		Interleaving &
		requested &
		Received a Downlink Data Notification Ack in SGW-C operational state \\
		\hline
		C14 & open5gs & SGW-C &
		Att $||$ Det &
		Nested &
		requested &
		Received a Session Modification response after session modification \\
		\hline
		C15 & open5gs & SGW-C &
		Att $||$ Det &
		Nested &
		requested &
		Received a Session Modification response after a session release \\
		\hline
		C16 & CC-A & MME &
		Att $||$ Det &
		Nested &
		- &
		Received a create session response after UE context release \\
		\hline
		C17 & CC-A & AMF &
		SR $||$ UEConRel &
		Nested &
		- &
		Received a Service Request response after UE context release \\
		\hline
		C18 & CC-B & AMF &
		Reg &
		Incomplete &
		- &
		Received no Create UE context response, causing a null pointer \\
		\hline
		P1 & free5gc & AUSF &
		Reg &
		Incomplete &
		requested &
		Received no GenerateAuthData response, causing a null pointer \\
		\hline
		P2 & free5gc & UDM &
		Reg &
		Incomplete &
		requested &
		Received no GenerateAuthData response, causing a null pointer \\
		\hline
	\end{tabular}
	
	\vspace{0.5ex}
	\begin{minipage}{\linewidth}
		\footnotesize
		\textbf{Note.} Abbreviations:
		C: Network Function \textbf{C}rash,
		P: Network Function \textbf{P}ersistently Unavailable,
		HO: \textbf{H}andover,
		Reg: \textbf{Reg}istration,
		DeReg: \textbf{De}-\textbf{reg}istration,
		Att: \textbf{Att}ach,
		Det: \textbf{Det}ach,
		SR: \textbf{S}ervice \textbf{R}equest,
		UEConRel: \textbf{UE} \textbf{Con}text \textbf{Rel}ease,
		CC-A/CC-B: Two \textbf{C}ommercial \textbf{C}ellular Core Networks.
	\end{minipage}
\end{table*}

\subsection{Identified Failures}
\label{subsec:identified-failures}
To address RQ1, we evaluate the generality of Kairos by deploying it on two commercial and two open-source LTE and 5G cellular core networks. 
As summarized in Table~\ref{tab:vulnerability-overview}, Kairos uncovers 20 new timing-induced interaction failures. 
In addition, Kairos successfully reproduces 34 additional issues.
These reproduced issues include previously reported crashes, confirmed bug reports, and known implementation issues documented by developers.
Among the new identified failures, 11 issues have been confirmed by developers and assigned CVE identifiers, demonstrating that the failures exposed by Kairos are reproducible in real cellular core network deployments. 
In particular, a subset of these failures were uncovered in commercial core network deployments, indicating that timing-induced interaction failures are not limited to open-source implementations. 
Commercial core network deployments are anonymized in this study to comply with responsible disclosure requirements.

By deploying Kairos in LTE and 5G deployments, we identify 5 failures (\textbf{C12-C16}) in LTE core networks and 15 failures (\textbf{C1-C11, C17-C18, P1-P2}) in 5G core networks.
We observe that more failures are identified in 5G core networks than in LTE.
This trend is consistent with the architectural evolution of 5G core networks, where control-plane functions are further decomposed into a larger set of network functions and interconnected through service-based interfaces. 
As a result, control-plane procedures are more frequently executed in parallel and involve a larger number of interactions across network functions. 
This increases the probability of timing-induced interaction failures.

Kairos constructs diverse UE behaviors to explore different combinations of control-plane procedures for identifying timing-induced interaction failures.
Among the 20 identified failures, 14 are related to registration behaviors (\textbf{C1-C11, C18, P1-P2}), while 5 are related to attach procedures (\textbf{C12-C16}).
This is mainly because registration and attach procedures involve long execution paths with multiple dependent stages and shared control-plane states, making them particularly sensitive to relative execution timing.
These procedures also span relatively long execution paths and include multiple dependent processing stages, which makes them more sensitive to execution timing.
Finally, failures can also be triggered by less frequent UE behaviors (\textbf{C17}), as observed in commercial core network~A. 
It indicates that timing-induced failures are not limited to a small set of core network procedures.

Kairos identifies timing-induced interaction failures in different core network functions, including AMF, MME, SMF, SGW-C, AUSF, and UDM.
We observe that most failures occur in access and mobility control functions (\textbf{C1-C9, C12, C16-C18}), the next are session and forwarding control functions (\textbf{C10-C11, C13-C15}), while only a small number of cases are observed in authentication and user data management functions (\textbf{P1-P2}).
This is likely because AMF and MME participate in the largest number of control-plane procedures and maintain the most complex UE-related states.
In contrast, SMF and SGW-C mainly handle session establishment, modification, and release, and are involved in a more limited set of control-plane procedures.
AUSF and UDM are involved in fewer control-plane procedures, which reduces their exposure to timing-induced failures.

Kairos identifies timing-induced interaction failures in both open-source and commercial cellular core networks.
We observe that the identified failures exhibit clear vendor-specific characteristics.
Such differences may be related to variations in architectural design and implementation choices across different core network implementations.
In addition, different implementation choices, including programming languages and development practices, may further influence how timing-induced failures are manifested.
For example, open5gs is primarily implemented in C, while free5gc is implemented in Go.
This observation suggests that timing-induced interaction failures are shaped not only by protocol logic, but also by implementation-level design decisions.
It further highlights the necessity of testing approaches that are agnostic to specific architectures and programming languages.

Taken together, the identified failures span different core network generations, diverse UE behaviors, all major control-plane network functions, and both open-source and commercial core networks. 
Kairos can systematically expose such failures across diverse cellular core network settings.
They represent a general class of failures that are prevalent in LTE and 5G core networks.
%-------------------------------------------------------------------------------

\subsection{Effectiveness Across Interaction Patterns}
To address RQ2, we evaluate how effectively Kairos can identify timing-induced interaction failures across different interaction patterns.
Kairos does not explicitly model or enforce specific interaction patterns like interleaving, nested, or incomplete interactions. 
Instead, it perturbs execution timing at network functions, which increases the likelihood of observing failures across different interaction patterns during realistic network executions.
In this context, the effectiveness of Kairos is reflected by its ability to systematically identify failures associated with interaction patterns.
As summarized in Table~\ref{tab:vulnerability-overview}, Kairos identifies failures associated with interleaving (\textbf{C1-C13}), nested (\textbf{C14-C17}), and incomplete (\textbf{C18, P1-P2}) interaction patterns across both LTE and 5G core networks.

\subsubsection{Interleaving Interaction Failures}
As shown in Table~\ref{tab:vulnerability-overview}, interleaving interactions account for the majority of the identified timing-induced failures.
Specifically, 13 out of the 20 newly discovered failures fall into the interleaving category, and interleaving interactions are also involved in 24 out of the 34 reproduced issues.
We identify two common failure mechanisms that are responsible for most of observed interleaving-related issues: \textbf{missing state transitions(C1, C3, C4, C6-C9, C11-C13)} and \textbf{missing processing context(C2, C5, C10)}.

\begin{figure}[t]
	\centering
	\includegraphics[width=1\columnwidth]{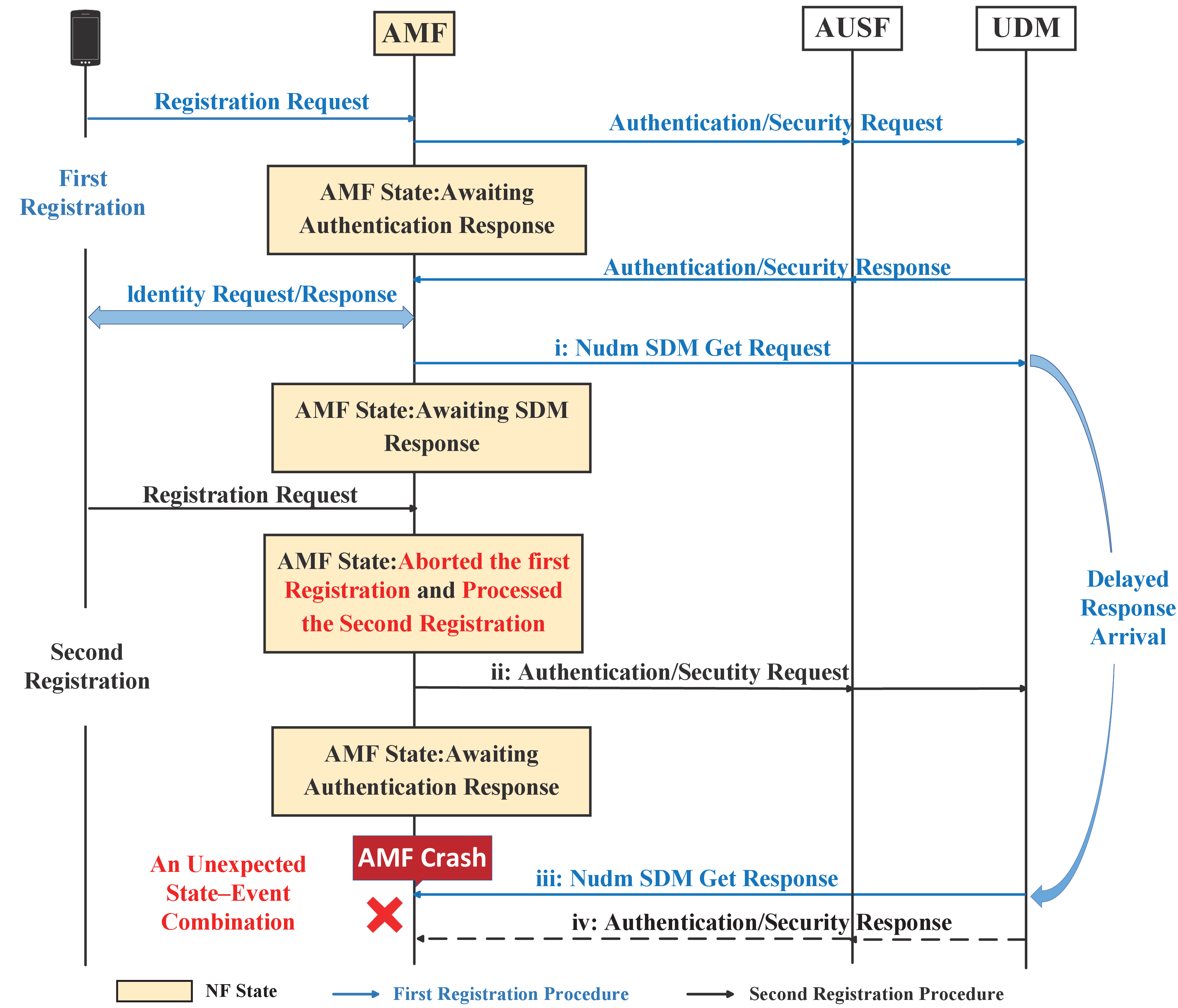}
	\caption{An interleaving interaction failure caused by two  registration procedures.}
	\label{fig:amf_crash}
\end{figure}

The first class of failures happens when a network function is unable to transition to the correct state after receiving a message, which is referred to as a \textbf{missing state transition}. 
In \textbf{C1}, as illustrated in Figure~\ref{fig:amf_crash}, we focus on interleaving interactions between two concurrent registration procedures.
The first registration procedure sends an SDM request(i) to the UDM and later receives the response(iii).
Before the SDM response arrives, the UE initiates a second registration procedure, which triggers an authentication request(ii) to the AUSF and receives the response(iv).
Due to network latency between UDM and AMF, the SDM response(iii) arrives late.
By that time, the internal state of AMF has already been updated by the authentication request(ii) to wait for authentication related response(iv).
When the SDM response(iii) is processed under this state, there is no valid state transition exists.
As a result, the state machine enters an abnormal execution path and eventually crashes the network function.

The second class of interleaving failures arises from \textbf{missing processing context}.
In~\textbf{C5}, the AMF receives an SDM response while operating in the initial-context-setup state.
Different from the previous case, the AMF is able to process SDM responses in this state.
However, due to the interleaving execution of concurrent procedures, the UE context associated with the previous procedure has already been released or overwritten when the delayed SDM response is handled.
As a result, although the state machine permits handling the response, the required processing context is no longer available, which eventually causes the AMF to crash.

%\noindent\textbf{Further Analysis.}
By perturbing execution timing, Kairos is able to observe diverse interleaving interaction failures in practice.
Our analysis shows that interleaving failures predominantly arise from concurrent executions of identical control-plane procedures, such as two registration procedures or two attach procedures.
These failures are most frequently observed in access and mobility management functions, particularly AMF and MME, which maintain complex UE related states.
As described in Section~\ref{sec:background}, 3GPP specifications define how to handle conflicting procedures, e.g., by aborting the previous procedure when a new one arrives.
However, the specifications do not further define how such procedure abortion should be implemented.
In practice, core networks typically realize procedure abortion by resetting UE related states and releasing the existing UE context before creating a new one.
As a result, UE state and context are mutually exclusive across concurrent instances of the same procedure.

\begin{figure}[t]
	\centering
	\includegraphics[width=1\columnwidth]{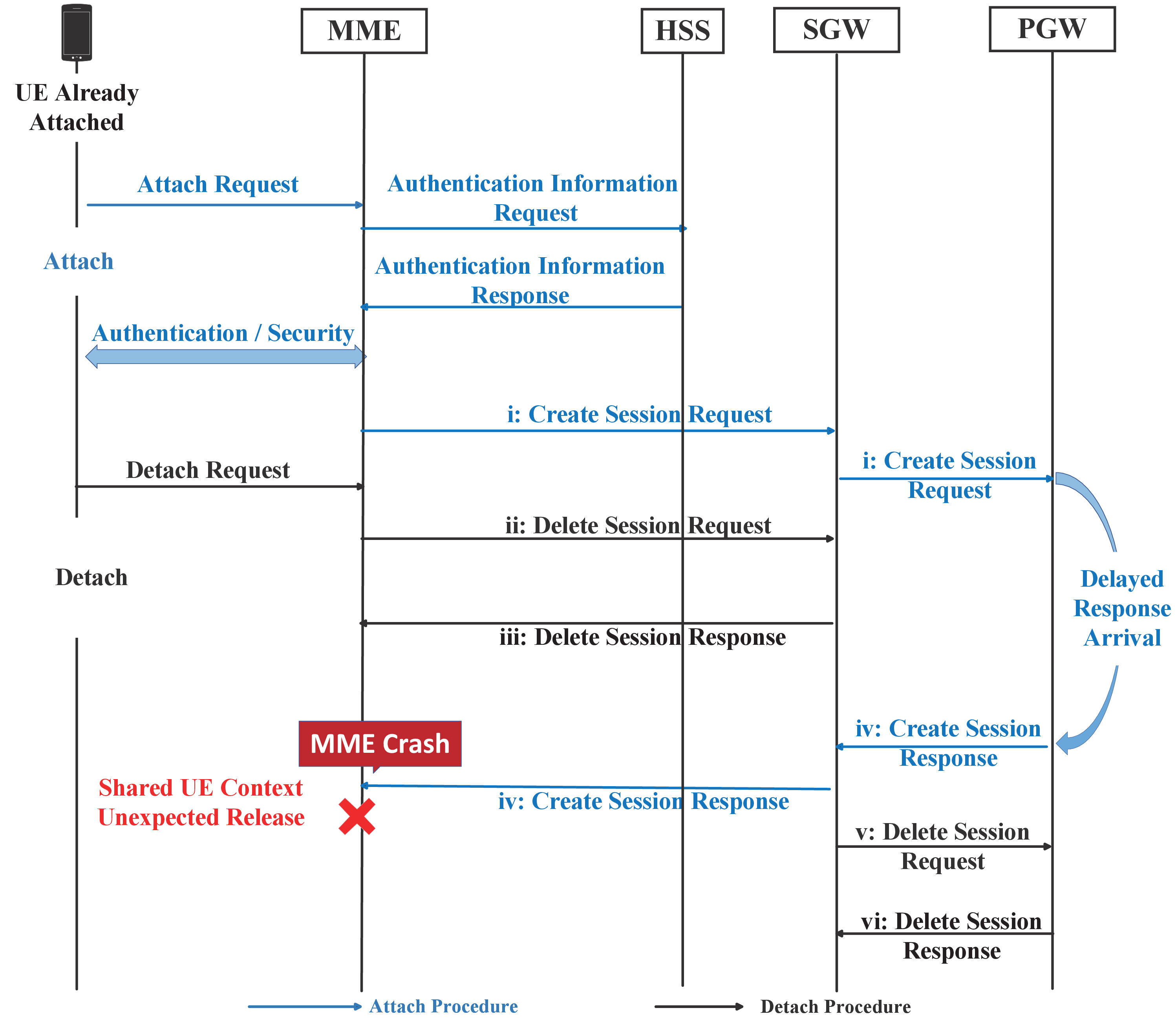}
	\caption{A nested interaction failure caused by the attach and detach procedures initiated by an attached UE.}
	\label{fig:mme_crash}
\end{figure}
\subsubsection{Nested Interaction Failures}
As summarized in Table~\ref{tab:vulnerability-overview}, Kairos identifies nested interactions in 4 of the 20 newly discovered failures(\textbf{C14-C17}).
Additionally, Kairos also identifies 7 nested interactions among the 34 reproduced issues.
Different from interleaving interaction failures, nested interaction failures arise when different procedures independently \textbf{modify or delete shared resources} without explicit coordination.

As shown in Figure~\ref{fig:mme_crash}, \textbf{C16} is a commercial MME nested interactions failures. 
Upon receiving the session delete request (ii) from MME, SGW first returns a session delete response (iii) to MME. Then, SGW asynchronously sends a session delete request (v) to PGW and receives the response (vi) from PGW.
After MME sends the create session request(i), it processes a complete Detach procedure(ii,iii).
Then MME receive the delayed create session response, which leads to an MME crash caused by the deletion of the shared UE context.

In summary, Kairos can identify nested interaction failures across core networks from both open-source and commercial vendors.
Unlike interleaving interaction failures, nested interaction failures often arise from asynchronous modifications or deletions of shared context.  
These failures typically involve two procedures: one procedure has more complex interactions, such as the attach procedure(i), and another has fewer interactions, such as detach(ii).  
In conclusion, nested interaction failures are more likely to occur when one of the procedures with NF delay, while the other process normally.
\begin{figure}[t]
	\centering
	\includegraphics[width=0.8\columnwidth]{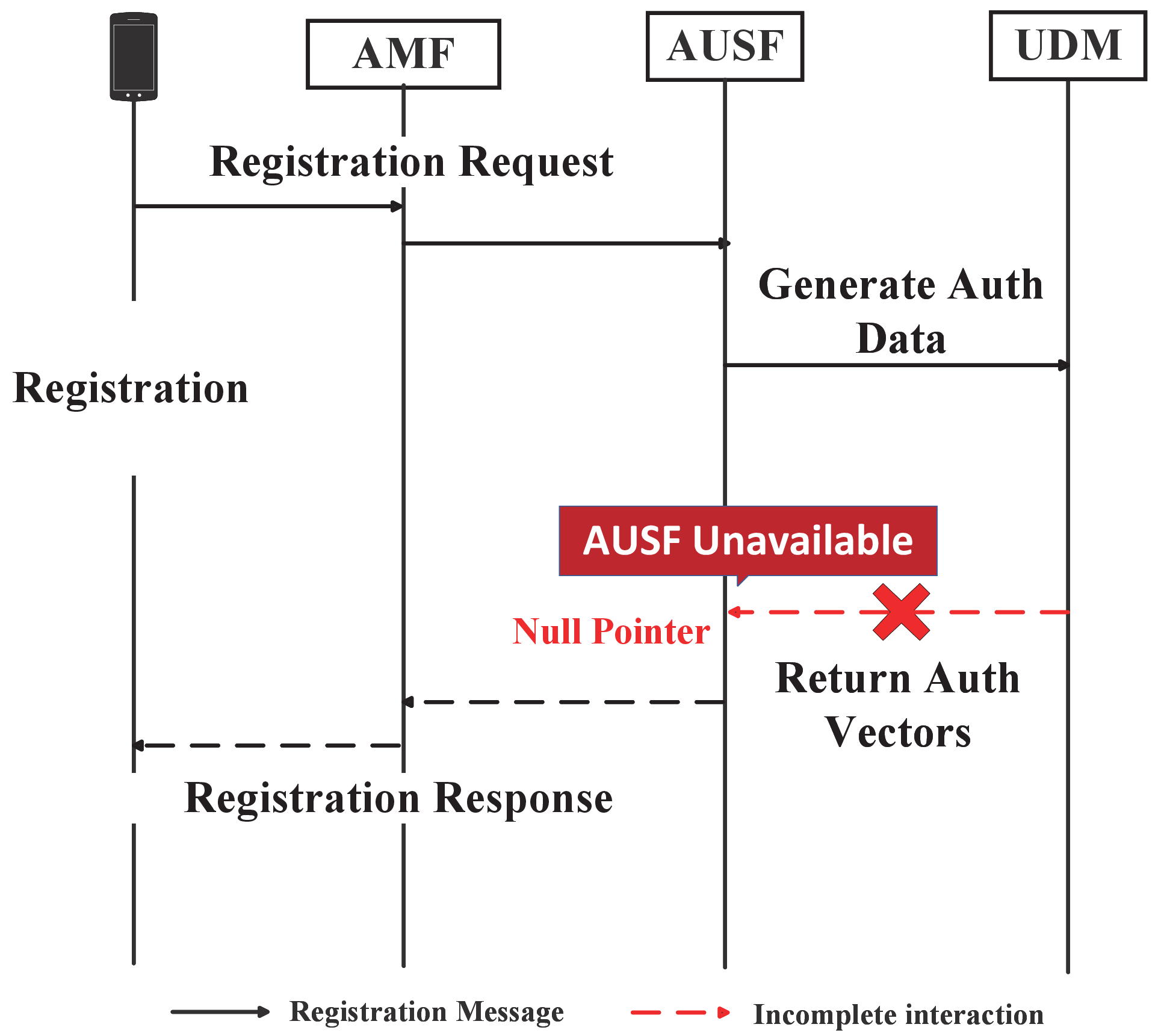}
	\caption{An incomplete interaction failure caused by return Auth vectors response fail.}
	\label{fig:ausf_incomplete}
\end{figure}
\subsubsection{Incomplete Interaction Failures}
In total, Kairos identifies incomplete interaction failures in 3 of the 20 newly discovered failures(\textbf{C18, P1-P2}) and 3 among the 34 reproduced issues.
Incomplete failures typically arise when network function procedures are not completed, resulting in \textbf{resources not being successfully created}.
As shown in Figure~\ref{fig:ausf_incomplete}, \textbf{P1} is a AUSF  persistently unavailable failures caused by an incomplete HTTP2 authentication vector response. 
This failure occurs because the system expects every response to return a result without validating the response.
As a result, incomplete interaction failures happen when procedures attempt to use a context that was not successfully created, regardless of the current or other procedures.

\begin{figure}[t]
	\centering
	\includegraphics[width=1\columnwidth]{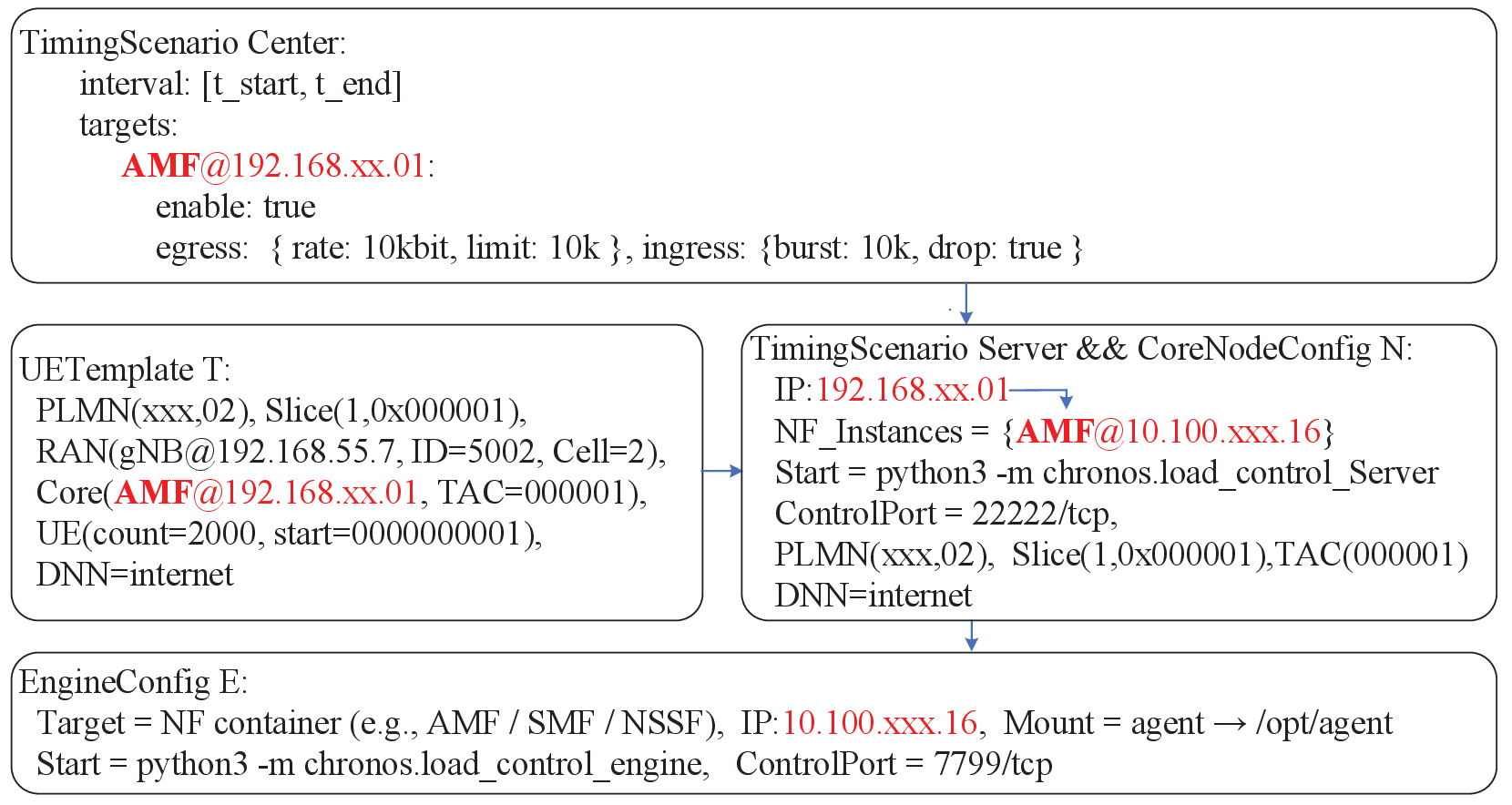}
	\caption{Operational view of Kairos configuration and deployment}
	\label{fig:config}
\end{figure}

\subsection{Efficiency and Deployment Overhead}
To address RQ3, we evaluate Kairos from the perspective of efficiency and deployment overhead in identifying timing-induced interaction failures. 
As the first systematic study of timing-induced interaction failures in LTE and 5G core networks, Kairos should not only efficiently identify timing-induced interaction failures, but also demonstrate that such identification can be achieved with low deployment and operational overhead in practice.
%Compared to existing works~\cite{al2024hermes, chen2023sherlock}, it significantly reduces protocol analysis and preprocessing time. 

\noindent\textbf{Efficiency.}~We first evaluate the efficiency of Kairos from two aspects: preparation overhead and failure discovery time.
Existing approaches typically require analysis of protocol specifications before testing, which can take several hours or more in practice~\cite{al2024hermes, chen2023sherlock}.
In the preparation phase, Kairos does not require analyzing 3GPP specifications or constructing malformed message test cases.
Instead, testing scenarios are configured by selecting and instantiating predefined UE config templates in the Kairos UE Behavior Generator.
As shown in Figure~\ref{fig:config}, UE templates are readily available and can be configured within a few minutes, enabling rapid setup for testing a target core network.

Kairos also demonstrates high efficiency in identifying timing-induced interaction failures.
As shown in Table~\ref{tab:failure-detection-time}, the majority of failures are exposed within 120 seconds,
with several failures being identified in less than 10 seconds.
Interleaving interaction failures are generally detected the fastest,
followed by incomplete interactions, while nested interaction failures typically require longer time to manifest.We further observe that failure detection is influenced not only by the injected timing perturbations,
but also by the frequency of UE behavior initiations.
This observation is consistent with prior studies on metastable failures~\cite{bronson2021metastable, huang2022metastable, isaacs2025analyzing},
where failures may only surface under specific combinations of timing and execution frequency.

\begin{table}[t]
	\centering
	\caption{Failure Detection Time Across Interaction Patterns}
	\label{tab:failure-detection-time}
	\setlength{\tabcolsep}{10pt}
	\scriptsize
	\begin{tabular}{lccccc}
		\hline
		\textbf{Interaction Pattern} &
		\textbf{10s} &
		\textbf{30s} &
		\textbf{60s} &
		\textbf{120s} &
		\textbf{120s+} \\
		\hline
		Interleaving & 5 & 5 & 2 & 1 & 0 \\
		Nested       & 0 & 0 & 0 & 1 & 3 \\
		Incomplete   & 0 & 0 & 2 & 0 & 1 \\
		\hline
	\end{tabular}
\end{table}

\noindent\textbf{Deployment Overhead.}~We further evaluate the deployment overhead of Kairos on two open-source and two commercial cellular core networks.
Kairos treats the target core network as a black box and does not require analyzing internal implementations or modifying core network code.
As shown in Figure~\ref{fig:config}, the Network Impairment Engine can be deployed within existing containers, making Kairos suitable for both single-node and large-scale deployments.
In practice, deploying Kairos only requires configuring the Scenario Orchestrator center and launching a lightweight Scenario Orchestrator server on the physical host of the core network.
Migration across different core network implementations is achieved by adjusting the orchestration configuration,
without changes to the testing logic or engine design.
Across all evaluated deployments, the complete setup process can be completed within 10 minutes, after which Kairos is ready to conduct testing.

\section{Related Works}
%-------------------------------------------------------------------------------
\noindent\textbf{Fuzzing.} A line of prior fuzzing work focuses on analyzing protocol specifications or inferred protocol models to generate test cases for security testing of user equipments \cite{kim2019touching, hussain2021noncompliance, chen2023sherlock}.
Hussain et al. propose Hermes which can automatically generate formal representations from natural language cellular specifications\cite{al2024hermes}. CORECRISIS implements a stateful black-box testing framework for 5G core network implementations, identified 7 categories of deviations from the technical specifications and 13 crashing vulnerabilities\cite{dong2025corecrisis}. Sun et al. construct a 5GC state machine and a grammar-aware signaling sequence mutation method, finding 22 security-critical vulnerabilities in 5G core networks\cite{sun20255gc}. Bianchi et al. focus on black-box fuzzing of the AMF and demonstrate its effectiveness on three open-source 5G core network implementations\cite{mancini2024amfuzz}.

\noindent\textbf{Formal analysis.}~Formal analysis has been widely applied to the security analysis of cellular control-plane protocols. Early work demonstrates systematic, property-driven adversarial analysis for 4G LTE procedures, laying the groundwork for formal reasoning about protocol state, message ordering, and adversarial capabilities\cite{hussain2018lteinspector}. With the advent of 5G, extensive symbolic analyses focus on authentication and key agreement protocols, identifying missing security goals, implicit trust assumptions, and session confusion issues in 5G-AKA and its variants\cite{basin2018formal, hussain20195greasoner, cremers2019component, wang2021privacy}. Beyond authentication, formal verification has been extended to other critical control-plane procedures, including handover protocols and EAP-based authentication, highlighting the dependence of security guarantees on precise assumptions about message ordering and protocol state evolution\cite{peltonen2021comprehensive, shi2025formal}. More recently, the service-based architecture of the 5G core has motivated formal analyses of access control and network function interactions, uncovering over-privilege and authorization flaws through model checking and compositional verification techniques\cite{akon2023formal,akon2025control,suarez2020formalization}. In contrast, our work targets time-induced failures by perturbing message timing and execution order, complementing prior formal analyses with a fuzzing-based exploration of execution-level vulnerabilities.

%-------------------------------------------------------------------------------
\section{Discussion}
%-------------------------------------------------------------------------------
%Kairos is not intended to exhaustively enumerate all timing-induced behaviors in cellular core networks.
%Instead, it provides a mechanism to amplify short-lived execution windows and make otherwise hard-to-observe control-plane interactions externally visible.
%While some of the observed failures could be avoided by enforcing strictly serialized execution of control-plane procedures, such designs are rarely adopted in practice.
%Modern cellular core networks are inherently distributed systems, where network functions must balance correctness with execution efficiency, and often make local decisions to terminate or advance procedures independently.
%Ensuring globally consistent termination semantics across multiple network functions requires explicit coordination, which is difficult to achieve under asynchronous execution and variable timing conditions.
%As a result, the failures exposed by Kairos should not be viewed as artifacts of specific implementations, but as manifestations of broader execution-level challenges introduced by distributed and performance-driven core network architectures.

%Discussions
\noindent\textbf{Low-Probability, High-Impact} characteristics of timing-induced interaction failures.
Timing-induced interaction failures arise when two well-formed control-plane interactions execute under specific timing.
As discussed in Section~\ref{subsec:Challenges and Design Rationales}, such failures typically occur within narrow timing windows and therefore exhibit a low probability under normal operating conditions. 
However, once triggered, their consequences are often disproportionate to their likelihood. 
As analyzed in Section~\ref{sec:taxonomy} and validated in Section~\ref{sec:Evaluation}, these failures disrupt the temporal consistency between control-plane interactions and their execution contexts, leading to corrupted per-UE state, broken procedure continuity, and in some cases, network function crashes.
This combination of low observability and high impact makes timing-induced interaction failures particularly difficult to uncover through conventional testing, while allowing them to manifest as abrupt and severe failures in operational core networks.
%-------------------------------------------------------------------------------

\noindent\textbf{Cascading Effects} caused by timing-induced interaction failures.
Although timing-induced interaction failures occur infrequently under normal conditions, their impact can be amplified in large-scale operational core networks, where stability mechanisms such as redundancy and failover are routinely activated.
In such environments, failures of individual network functions commonly trigger traffic redirection and control-plane recovery procedures to maintain service continuity.
As control-plane traffic is redistributed, subsequent interactions are executed under increased concurrency and load, which increases the likelihood of timing-induced interaction failures.
Such failures can trigger additional traffic redirection, increasing the load on other network functions and, in turn, raising the likelihood of further timing-induced interaction failures.
As a result, timing-induced interaction failures can give rise to cascading effects that propagate across network functions and undermine overall control-plane stability.

\noindent\textbf{Necessity of Systematic Fuzzing} for timing-induced interaction failures.
Because timing-induced interaction failures are both hard to detect and potentially severe in impact, systematic pre-deployment testing becomes necessary.
Kairos is a lightweight testing framework that exposes timing-induced interaction failures in practice, but can not discover all failures.
Further techniques are still required to more systematically explore timing-induced interaction failures.

\section{Conclusion}
In this paper, we conduct a systematic study of timing-induced interaction failures in LTE and 5G cellular core networks.
We first establish a taxonomy of control-plane interaction patterns and analyze the failure modes associated with each pattern.
Then, we design and implement Kairos, a lightweight testing framework that exposes timing-induced interaction failures without analyzing cellular standard documents.
We evaluate Kairos on two open-source and two commercial LTE and 5G core network implementations.
Through this evaluation, we uncover 20 previously unknown vulnerabilities and reproduce 34 existing issues across different network functions, UE behaviors, and core network deployments.
Our results show that timing-induced interaction failures are prevalent in both LTE and 5G core networks and are not limited to specific implementations or vendors.
These findings indicate that timing-induced interaction failures should be explicitly considered in the design, implementation, and evolution of cellular core network specifications.

%-------------------------------------------------------------------------------

%-------------------------------------------------------------------------------
%\section*{Acknowledgments}
%%-------------------------------------------------------------------------------
%
%The USENIX latex style is old and very tired, which is why
%there's no \textbackslash{}acks command for you to use when
%acknowledging. Sorry.
%
%\textbf{Do not include any acknowledgements in your submission which may deanonymize you (e.g., because of specific affiliations or grants you acknowledge)}

%-------------------------------------------------------------------------------
% optional clearing of the page
%\cleardoublepage
\appendix
\section*{Ethical Considerations}
The core networks validated in this study include two open-source and two commercial core networks with explicit permission, all of which are deployed in a controlled laboratory environment. As summarized in Table ~\ref{tab:vulnerability-overview}, we have disclosed 20 vulnerabilities to the core network vendors and developers, 11 of which have already been fixed and assigned CVE identifiers. 
Furthermore, because some of these failures can be triggered through normal UE interactions under specific conditions, we anonymize the affected commercial core network vendors to reduce the risk of misuse.
% optional clearing of the page
%\cleardoublepage

\section*{Open Science}
We make the Kairos testing framework available to support the evaluation and reuse of our work. 
The artifact includes the implementation of Kairos and representative configurations used in our experiments. 
All materials are provided through an anonymous repository and can be accessed at the following link: \url{https://anonymous.4open.science/r/Kairos-FC84/Readme.md}.
%\textbf{Within up to one page, this appendix must list all artifacts necessary to evaluate the contribution of the paper and make clear how the review committees can access each artifact. This appendix must have exactly this title, otherwise you will risk desk rejection. }

% optional clearing of the page
\cleardoublepage
\bibliographystyle{unsrt}
\bibliography{\jobname}

\begin{thebibliography}{10}

\bibitem{3gpp:ts23401}
{3GPP}.
\newblock {General Packet Radio Service (GPRS) Enhancements for Evolved
  Universal Terrestrial Radio Access Network (E-UTRAN) Access}.
\newblock {Technical Specification (TS) 23.401}, 2022.
\newblock {Version18.0.0}.

\bibitem{3gpp:ts23501}
{3GPP}.
\newblock {System Architecture for the 5G System}.
\newblock {Technical Specification (TS) 23.501}, 2025.
\newblock {Version 18.12.0}.

\bibitem{3gpp:ts33501}
{3GPP}.
\newblock {Security Architecture and Procedures for 5G System}.
\newblock {Technical Specification (TS) 33.501}, 2025.
\newblock {Version 18.10.0}.

\bibitem{3gpp:ts23502}
{3GPP}.
\newblock {Procedures for the 5G System (5GS)}.
\newblock {Technical Specification (TS) 23.502}, 2025.
\newblock {Version 18.12.0}.

\bibitem{3gpp:ts24501-r18}
{3GPP}.
\newblock {Non-Access-Stratum (NAS) Protocol for 5G System (5GS); Stage~3}.
\newblock {Technical Specification (TS) 24.501}, 2025.
\newblock {Version 18.13.1}.

\bibitem{3gpp:ts33512}
{3GPP}.
\newblock {5G Security Assurance Specification (SCAS); Access and Mobility
  management Function (AMF)}.
\newblock {Technical Specification (TS) 33.512}, 2023.
\newblock {Version 18.0.0}.

\bibitem{open5gs}
{Open5GS}.
\newblock {Open Source 5G Core Network}.
\newblock \url{https://open5gs.org}.

\bibitem{free5gc}
{Free5GC}.
\newblock {Open Source 5G Core Network}.
\newblock \url{https://www.free5gc.org}.

\bibitem{srsran}
{SRSRAN Project}.
\newblock {Open Source RAN and Core Network Software}.
\newblock \url{https://www.srsran.com}.

\bibitem{ueransim}
{UERANSIM}.
\newblock {Open Source 5G UE and RAN Simulator}.
\newblock \url{https://github.com/aligungr/UERANSIM}.

\bibitem{sekigawa2022toward}
Shu Sekigawa, Chikara Sasaki, and Atsushi Tagami.
\newblock Toward a cloud-native telecom infrastructure: Analysis and
  evaluations of kubernetes networking.
\newblock In {\em 2022 IEEE Globecom Workshops (GC Wkshps)}, pages 838--843.
  IEEE, 2022.

\bibitem{mudvari2022exploring}
Akrit Mudvari, Nikos Makris, and Leandros Tassiulas.
\newblock Exploring ml methods for dynamic scaling of beyond 5g cloud-native
  rans.
\newblock In {\em ICC 2022-IEEE International Conference on Communications},
  pages 2284--2289. IEEE, 2022.

\bibitem{dalgitsis2024cloud}
Michail Dalgitsis, Nicola Cadenelli, Maria~A Serrano, Nikolaos Bartzoudis, Luis
  Alonso, and Angelos Antonopoulos.
\newblock Cloud-native orchestration framework for network slice federation
  across administrative domains in 5g/6g mobile networks.
\newblock {\em IEEE Transactions on Vehicular Technology}, 73(7):9306--9319,
  2024.

\bibitem{sun20255gc}
Yu~Sun, Xinyu Liu, Qian Sun, Jiaming Wang, Lin Tian, and Jianwei Liu.
\newblock 5gc-fuzz: Finding deep stateful vulnerabilities in 5g core network
  with black-box fuzzing.
\newblock In {\em IEEE INFOCOM 2025-IEEE Conference on Computer
  Communications}, pages 1--10. IEEE, 2025.

\bibitem{dong2025corecrisis}
Yilu Dong, Tianchang Yang, Abdullah Al~Ishtiaq, Syed Md~Mukit Rashid, Ali
  Ranjbar, Kai Tu, Tianwei Wu, Md~Sultan Mahmud, and Syed~Rafiul Hussain.
\newblock Corecrisis:threat-guided and context-aware iterative learning and
  fuzzing of 5g core networks.
\newblock In {\em 34th USENIX Security Symposium (USENIX Security 25)}, pages
  5287--5306, 2025.

\bibitem{son2025citesting}
Mincheol Son, Kwangmin Kim, Beomseok Oh, CheolJun Park, and Yongdae Kim.
\newblock Citesting: Systematic testing of context integrity violations in lte
  core networks.
\newblock In {\em Proceedings of the 2025 ACM SIGSAC Conference on Computer and
  Communications Security}, pages 276--290, 2025.

\bibitem{al2024hermes}
Abdullah Al~Ishtiaq, Sarkar Snigdha~Sarathi Das, Syed Md~Mukit Rashid, Ali
  Ranjbar, Kai Tu, Tianwei Wu, Zhezheng Song, Weixuan Wang, Mujtahid Akon, Rui
  Zhang, et~al.
\newblock Hermes: Unlocking security analysis of cellular network protocols by
  synthesizing finite state machines from natural language specifications.
\newblock In {\em 33rd USENIX Security Symposium (USENIX Security 24)}, pages
  4445--4462, 2024.

\bibitem{rahman2024cellularlint}
Mirza~Masfiqur Rahman, Imtiaz Karim, and Elisa Bertino.
\newblock Cellularlint: A systematic approach to identify inconsistent behavior
  in cellular network specifications.
\newblock In {\em 33rd USENIX Security Symposium (USENIX Security 24)}, pages
  5215--5232, 2024.

\bibitem{wei2025unleashing}
Haiyang Wei, Ligeng Chen, Zhengjie Du, Yuhan Wu, Haohui Huang, Yue Liu, Guang
  Cheng, Fengyuan Xu, Linzhang Wang, and Bing Mao.
\newblock Unleashing the power of llm to infer state machine from the protocol
  implementation.
\newblock In {\em 2025 IEEE/ACM 33rd International Symposium on Quality of
  Service (IWQoS)}, pages 1--10. IEEE, 2025.

\bibitem{saifuzzaman2025dissecting}
Munshi Saifuzzaman, Ke~Xie, Tian Xie, Xiao Zhang, and Xinyu Lei.
\newblock Dissecting privacy-exposing identifiers in 5g/4g networks.
\newblock In {\em 2025 IEEE Conference on Dependable and Secure Computing
  (DSC)}, pages 1--9. IEEE, 2025.

\bibitem{klischies2023instructions}
Daniel Klischies, Moritz Schloegel, Tobias Scharnowski, Mikhail Bogodukhov,
  David Rupprecht, and Veelasha Moonsamy.
\newblock Instructions unclear: undefined behaviour in cellular network
  specifications.
\newblock In {\em 32nd USENIX Security Symposium (USENIX Security 23)}, pages
  3475--3492, 2023.

\bibitem{hussain2021noncompliance}
Syed~Rafiul Hussain, Imtiaz Karim, Abdullah~Al Ishtiaq, Omar Chowdhury, and
  Elisa Bertino.
\newblock Noncompliance as deviant behavior: An automated black-box
  noncompliance checker for 4g lte cellular devices.
\newblock In {\em Proceedings of the 2021 ACM SIGSAC Conference on Computer and
  Communications Security}, pages 1082--1099, 2021.

\bibitem{kim2019touching}
Hongil Kim, Jiho Lee, Eunkyu Lee, and Yongdae Kim.
\newblock Touching the untouchables: Dynamic security analysis of the lte
  control plane.
\newblock In {\em 2019 IEEE Symposium on Security and Privacy (SP)}, pages
  1153--1168. IEEE, 2019.

\bibitem{chen2023sherlock}
Yi~Chen, Di~Tang, Yepeng Yao, Mingming Zha, XiaoFeng Wang, Xiaozhong Liu, Haixu
  Tang, and Baoxu Liu.
\newblock Sherlock on specs: Building lte conformance tests through automated
  reasoning.
\newblock In {\em 32nd USENIX Security Symposium (USENIX Security 23)}, pages
  3529--3545, 2023.

\bibitem{he2022intelligent}
Fengjiao He, Wenchuan Yang, Baojiang Cui, and Jia Cui.
\newblock Intelligent fuzzing algorithm for 5g nas protocol based on predefined
  rules.
\newblock In {\em 2022 International Conference on Computer Communications and
  Networks (ICCCN)}, pages 1--7. IEEE, 2022.

\bibitem{wang2023nlp}
Zhuzhu Wang and Ying Wang.
\newblock Nlp-based cross-layer 5g vulnerabilities detection via fuzzing
  generated run-time profiling.
\newblock In {\em 2023 IEEE 12th International Conference on Cloud Networking
  (CloudNet)}, pages 194--202. IEEE, 2023.

\bibitem{bronson2021metastable}
Nathan Bronson, Abutalib Aghayev, Aleksey Charapko, and Timothy Zhu.
\newblock Metastable failures in distributed systems.
\newblock In {\em Proceedings of the Workshop on Hot Topics in Operating
  Systems}, pages 221--227, 2021.

\bibitem{huang2022metastable}
Lexiang Huang, Matthew Magnusson, Abishek~Bangalore Muralikrishna, Salman
  Estyak, Rebecca Isaacs, Abutalib Aghayev, Timothy Zhu, and Aleksey Charapko.
\newblock Metastable failures in the wild.
\newblock In {\em 16th USENIX Symposium on Operating Systems Design and
  Implementation (OSDI 22)}, pages 73--90, 2022.

\bibitem{isaacs2025analyzing}
Rebecca Isaacs, Peter Alvaro, Rupak Majumdar, Kiran Kumar, Muniswamy Reddy,
  Mahmoud Salamati, and Sadegh Soudjani.
\newblock Analyzing metastable failures.
\newblock In {\em Proceedings of the 2025 Workshop on Hot Topics in Operating
  Systems}, pages 172--178, 2025.

\bibitem{mancini2024amfuzz}
Francesco Mancini, Sara Da~Canal, and Giuseppe Bianchi.
\newblock Amfuzz: Black-box fuzzing of 5g core networks.
\newblock In {\em 2024 19th Wireless On-Demand Network Systems and Services
  Conference (WONS)}, pages 17--24. IEEE, 2024.

\bibitem{hussain2018lteinspector}
Syed Hussain, Omar Chowdhury, Shagufta Mehnaz, and Elisa Bertino.
\newblock Lteinspector: A systematic approach for adversarial testing of 4g
  lte.
\newblock In {\em Network and Distributed Systems Security (NDSS) Symposium
  2018}, 2018.

\bibitem{basin2018formal}
David Basin, Jannik Dreier, Lucca Hirschi, Sa{\v{s}}a Radomirovic, Ralf Sasse,
  and Vincent Stettler.
\newblock A formal analysis of 5g authentication.
\newblock In {\em Proceedings of the 2018 ACM SIGSAC conference on computer and
  communications security}, pages 1383--1396, 2018.

\bibitem{hussain20195greasoner}
Syed~Rafiul Hussain, Mitziu Echeverria, Imtiaz Karim, Omar Chowdhury, and Elisa
  Bertino.
\newblock 5greasoner: A property-directed security and privacy analysis
  framework for 5g cellular network protocol.
\newblock In {\em Proceedings of the 2019 ACM SIGSAC Conference on Computer and
  Communications Security}, pages 669--684, 2019.

\bibitem{cremers2019component}
Cas Cremers and Martin Dehnel-Wild.
\newblock Component-based formal analysis of 5g-aka: Channel assumptions and
  session confusion.
\newblock In {\em Network and Distributed System Security Symposium (NDSS)}.
  Internet Society, 2019.

\bibitem{wang2021privacy}
Yuchen Wang, Zhenfeng Zhang, and Yongquan Xie.
\newblock Privacy-preserving and standard-compatible aka protocol for 5g.
\newblock In {\em 30th USENIX security symposium (USENIX security 21)}, pages
  3595--3612, 2021.

\bibitem{peltonen2021comprehensive}
Aleksi Peltonen, Ralf Sasse, and David Basin.
\newblock A comprehensive formal analysis of 5g handover.
\newblock In {\em Proceedings of the 14th ACM conference on security and
  privacy in wireless and mobile networks}, pages 1--12, 2021.

\bibitem{shi2025formal}
Min Shi, Jing Chen, Zhuangzhuang Ma, Kun He, Meng Jia, and Ruiying Du.
\newblock A formal analysis of 5g eap-tls protocol.
\newblock {\em IEEE Transactions on Networking}, 2025.

\bibitem{akon2023formal}
Mujtahid Akon, Tianchang Yang, Yilu Dong, and Syed~Rafiul Hussain.
\newblock Formal analysis of access control mechanism of 5g core network.
\newblock In {\em Proceedings of the 2023 ACM SIGSAC conference on computer and
  communications security}, pages 666--680, 2023.

\bibitem{akon2025control}
Mujtahid Akon, Md~Toufikuzzaman, and Syed~Rafiul Hussain.
\newblock From control to chaos: A comprehensive formal analysis of 5g's access
  control.
\newblock In {\em 2025 IEEE Symposium on Security and Privacy (SP)}, pages
  1081--1100. IEEE, 2025.

\bibitem{suarez2020formalization}
Luis Su{\'a}rez, David Espes, Fr{\'e}d{\'e}ric Cuppens, Philippe Bertin,
  Cao-Thanh Phan, and Philippe Le~Parc.
\newblock Formalization of a security access control model for the 5g system.
\newblock In {\em 2020 11th International Conference on Network of the Future
  (NoF)}, pages 150--158. IEEE, 2020.

\end{thebibliography}

%%%%%%%%%%%%%%%%%%%%%%%%%%%%%%%%%%%%%%%%%%%%%%%%%%%%%%%%%%%%%%%%%%%%%%%%%%%%%%%%
\end{document}